\documentclass[aps,prb,twocolumn,superscriptaddress,showpacs]{revtex4}
\usepackage{graphicx}

\begin{document}

\title{The boson-fermion model: An exact diagonalization study}

\author{M.~Cuoco}
\affiliation{Unit\`a I.N.F.M. di Salerno, Dipartimento di Fisica
``E. R. Caianiello'', Universit\`a di Salerno, I-84081 Baronissi
(Salerno), Italy}
\affiliation{Centre de Recherches sur les Tr\`es
Basses Temp\'eratures associ\'e \`a l'Universit\'e Joseph Fourier,
C.N.R.S., BP 166, 38042 Grenoble-C\'edex 9, France}

\author{C.~Noce}
\affiliation{Unit\`a I.N.F.M. di Salerno, Dipartimento di Fisica
``E. R. Caianiello'', Universit\`a di Salerno, I-84081 Baronissi
(Salerno), Italy} \affiliation{Unit\`a I.N.F.M. di Salerno -
Coherentia}

\author{J.~Ranninger}
\affiliation{Centre de Recherches sur les Tr\`es Basses
Temp\'eratures associ\'e \`a l'Universit\'e Joseph Fourier,
C.N.R.S., BP 166, 38042 Grenoble-C\'edex 9, France}

\author{A.~Romano}
\affiliation{Unit\`a I.N.F.M. di Salerno, Dipartimento di Fisica
``E. R. Caianiello'', Universit\`a di Salerno, I-84081 Baronissi
(Salerno), Italy}

\begin{abstract}
The main features of a generic boson-fermion scenario for electron
pairing in a many-body correlated fermionic system are: i) a
cross-over from a poor metal to an insulator and finally a
superconductor as the temperature decreases, ii) the build-up of a
finite amplitude of local electron pairing below a certain
temperature $T^*$, followed by the onset of long-range phase
correlations among electron pairs below a second characteristic
temperature $T_{\phi}$, iii) the opening of a pseudogap in the DOS
of the electrons below $T^*$, rendering these electrons poorer and
poorer quasi-particles as the temperature decreases, with the
electron transport becoming ensured by electron pairs rather than
by individual electrons. A number of these features have been so
far obtained on the basis of different many-body techniques, all
of which have their built-in shortcomings in the intermediate
coupling regime, which is of interest here. In order to
substantiate these features, we investigate them on the basis of
an exact diagonalization study on rings up to eight sites.
Particular emphasis has been put on the possibility of having
persistent currents in mesoscopic rings tracking the change-over
from single- to two-particle transport as the temperature
decreases and the superconducting state is approached.
\end{abstract}
\pacs{74.20.-z, 74.78.Na, 74.20.Mn} \maketitle

\section{Introduction}

A great interest has recently been devoted in condensed matter
theory to models for interacting boson-fermion systems. Among the
topics to which these models apply, we recall
the hole pairing in semiconductors \cite{Mysyrowicz-96}, the
isospin singlet pairing in nuclear matter \cite{Schnell-99},
$d$-wave hole and antiferromagnetic triplet pairing in the
positive-$U$ Hubbard system \cite{Auerbach-02} (and possibly also
in the t-J model), and entangled atoms in squeezed states in
molecular Bose-Einstein condensation in traps \cite{Yurovski-02}.

In the present paper we consider a boson-fermion model (BFM) which
captures a number of basic physical properties of strongly
interacting many-body systems where resonant pair states of
bosonic nature form inside a reservoir of fermions. This BFM was
originally conceived as a possible alternative and extension to
the idea of bipolaronic superconductivity \cite{Alexandrov-81a} -
a scenario for systems with extremely strong electron-phonon
coupling where all the electrons exist in form of locally bound
pairs of small polarons. Since in such a case of extreme
antiadiabaticity the mobility of bipolarons is expected to be
vanishingly small, they have to be considered, for all intents and
purposes, as remaining localized rather than condensing into a
superfluid phase. An alternative to this situation was to consider
a case where the electron-phonon coupling is of intermediate
strength, such that the energy of locally bound {\it isolated}
bipolarons is comparable with the gain in energy of itinerant
electrons. This leads to a scenario where two-electron states
fluctuate between pairs of itinerant single-electron states close
to the Fermi level and localized bipolaron states which pin this
Fermi level. A detailed discussion on such a polaron-induced
scenario for this BFM has been given recently \cite{Ranninger-02}.
Treating the bipolarons as bosonic entities which commute with the
fermionic electrons, the Hamiltonian for such a system is given by
\begin{eqnarray}
H & = & (D-\mu )\sum _{i\sigma }c_{i\sigma }^{+}c_{i\sigma }
 + (\Delta _{B}-2\mu )\sum _{i}\left(\rho_{i}^{z}+
\frac{1}{2}\right)\nonumber\\
& &\!\!\!\! + \,t\sum _{i\neq j,\: \sigma }c_{i\sigma
}^{+}c_{j\sigma } + g\sum _{i}\left( \rho^{+}_{i}c_{i\downarrow
}c_{i\uparrow }+ \rho^{-}_{i}c_{i\uparrow }^{+}c_{i\downarrow
}^{+}\right) \; .
\end{eqnarray}
The bipolarons are treated here as on-site entities having
hard-core boson features and thus are represented by pseudo-spin
operators $\{\rho_i^+,\rho_i^-,\rho_i^z\}$. $c^+_{i\sigma}$ and
$c_{i\sigma}$ denote the creation and annihilation operators of
the itinerant electrons and $g$ is the strength of the
boson-fermion charge exchange interaction. The hopping integral
for the itinerant electrons is given by $t$ with a half bandwidth
equal to $D = zt$, $z$ denoting the coordination number of the
underlying lattice. The energy level of the hard-core bosons is
denoted by $\Delta_B$. A common chemical potential $\mu$ ensures
the conservation of the total charge $n_{tot}=n_{F\uparrow} +
n_{F\downarrow} +2 n_B$ ($n_B$ and $n_{F\uparrow,\downarrow}$
denote the total number of the hard core-bosons and of the
electrons with up and down spin states). This model has been
studied by a number of different many-body techniques, such as
self-consistent diagrammatic studies \cite{Ranninger-95a},
dynamical mean-field theory \cite{Robin-98} and renormalization
group techniques \cite{Domanski-01}. The main message of all these
studies was that this model contained the following basic physical
properties:

{\it Upon decreasing the temperature below a certain
characteristic $T^*$, local electron pairing (not electron pair
states!) begin to form and exist over a finite time as well as
space interval. As a consequence of that, a pseudogap centered
around the Fermi level opens up in the density of states (DOS) of
the itinerant electrons. Electrical transport is being controlled
by relatively well defined single-electron states above $T^*$,
while below $T^*$ the electrons get heavily damped. Upon further
decreasing the temperature,  the electron pairs become better and
better defined quasi-particles and ensure the transport,
eventually leading to a superconducting state controlled by phase
fluctuations of the electron pairs. The resistivity, just before
entering  this superconducting state, resembles that of an
insulator, since it is determined by the electrons with their
pseudogap features.}

In the present paper, we examine all these features within an
exact diagonalization scheme, evaluating the static and dynamical
properties of the BFM, by means of the Lanczos procedure extended
to finite temperature\cite{Jaklic-94}. This technique has been
tested and successfully applied on different many-body systems of
strongly correlated electrons\cite{Jaklic-00}, and allows to
determine the temperature dependence of various physical
quantities, which, combined with dynamical properties, gives a
complete information about the energy spectrum and excitations of
the model under examination. In particular, within this scheme of
computation the trace of thermodynamic expectation values is
obtained by means of Monte Carlo sampling, while for the
determination of the relevant matrix elements the Lanczos method
is used.

We restrict ourselves to the symmetric case, where the bosonic
level lies in the middle of the electronic band, i.e.,
$\Delta_B=0$. For a different position of the bosonic level the
physics is qualitatively the same, the relevant parameter in this
scenario being the concentration of the bosons. Due to their
hard-core nature, they induce correlation effects which lead to
qualitative changes in the physical properties as we go from low
($\bar{n}_B \geq 0$) to high ($\bar{n}_B \simeq 0.5$) boson
concentrations.

Our local phase space being made up of eight states, a large
computational memory is required. This limits the present study to
maximally eight sites systems if the evaluation of the static and
dynamical quantities does not involve transitions between
subspaces with a different number of fermions and bosons.
Moreover, for the study of the spectral functions for the
fermions, together with the previous constraint we have to
consider a situation such that the discrete energy spectrum of our
finite size system has contributions as close as possible to the
Fermi level, in order to track the low energy features of interest
here. Having the Fermi level lie in the center of the fermionic
band (in order to avoid effects coming from any particular band
structure), the maximal size of the system which we could consider
is restricted to four sites rings allowing any concentration of
fermions and bosons. As far as concerns the boson spectral
function, we can investigate up to six sites rings since the
constraint on the physical description of the fermionic properties
does not apply.

\begin{figure}
\begin{center}
\includegraphics[width=3.4in,height=3.4in]{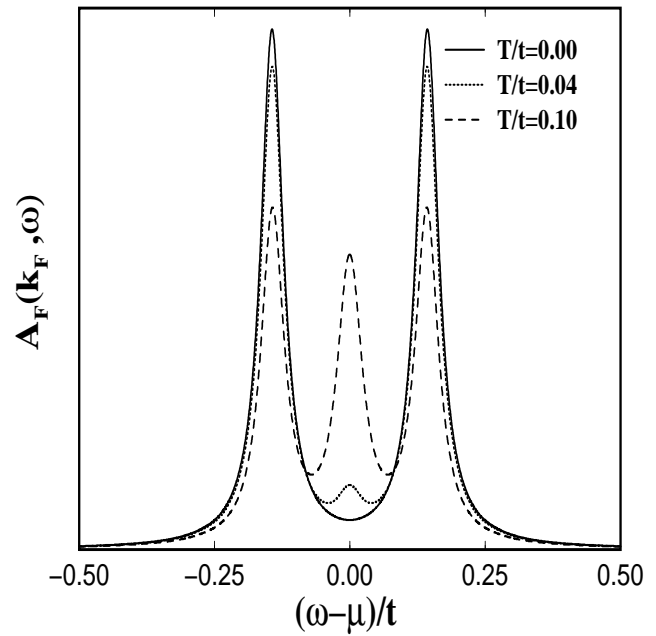}
\end{center}
\caption{\label{AF-T} Variation with temperature of the fermionic
spectral function at the Fermi vector, evaluated for a 4-site ring
with $\Delta_B=0$, $g/t=0.2$ and $n_{tot}=8$ ($n_B=2$).}
\end{figure}
\begin{figure}
\begin{center}
\includegraphics[width=3.4in,height=4.0in]{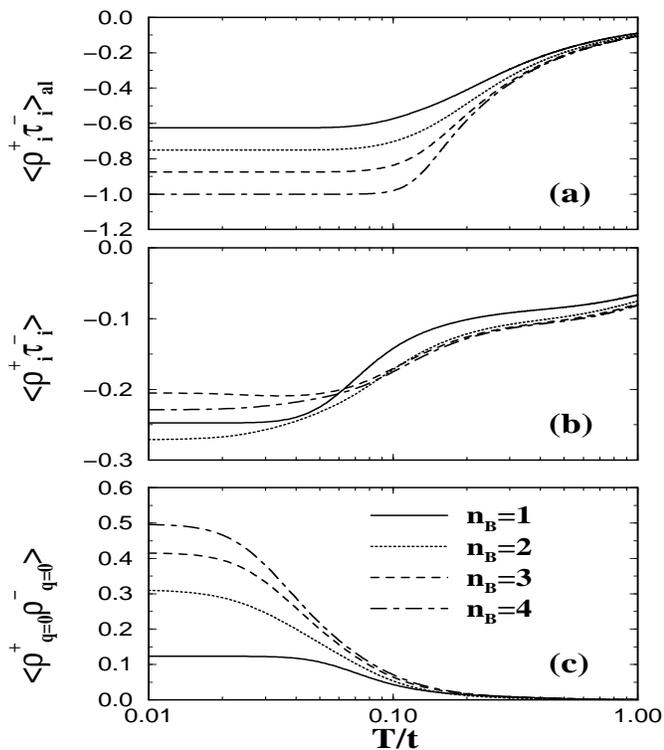}
\end{center}
\caption{\label{CF} Temperature dependence of the on-site
correlation function $\langle\rho^+_i \tau^-_i \rangle$, with
$\tau^-_i =c_{i\downarrow} c_{i\uparrow}$, evaluated for a single
site in the atomic limit (Fig.~1a). Comparison of this local
correlation function (Fig.~1b) and $\langle\rho^+_{q=0}
\rho^-_{q=0}\rangle$ (Fig.~1c), measuring the long-range phase
coherence of the hard-core bosons, evaluated for an 8-site system
for different boson concentrations $n_B$. The value of the
boson-fermion coupling is $g/t=0.4$.}
\end{figure}
\begin{figure}
\begin{center}
\includegraphics[width=3.4in,height=3.4in]{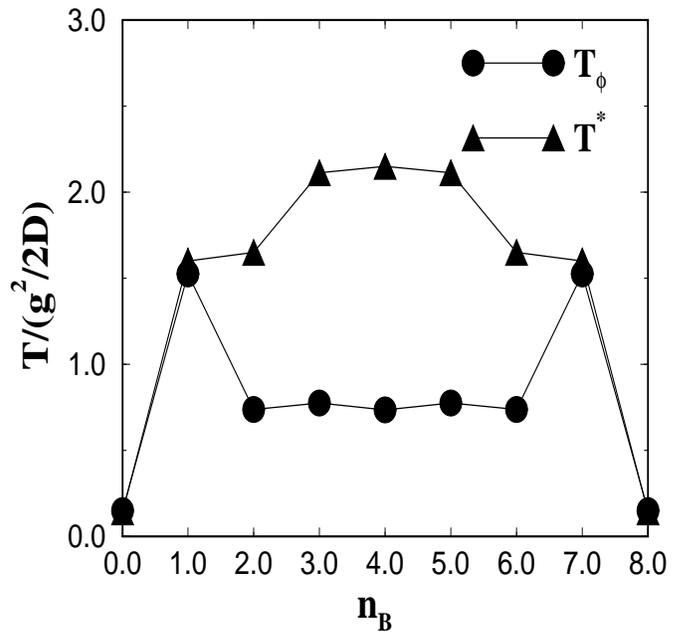}
\end{center}
\caption{\label{Tc} Variation of $T^*$ and $T_{\phi}$ as a
function of $n_B$ for an 8-site ring with $g/t=0.4$ and
$n_{tot}=n_B+8$.}
\end{figure}

The relevant correlation functions determining the onset of local
electron pairing and spatial phase correlations are given by
$\langle\rho^+_i c_{i\downarrow} c_{i\uparrow} + h.c.\rangle$ and
$\langle\rho^+_i \rho_j^- \rangle$ and will be discussed in
Section II. In Section III we examine the spectral properties of
the hard-core bosons and compare them with the excitations of an
equivalent XY model in the presence of a transverse field. The
change-over from single-electron to electron pair transport will
be discussed in Section IV on the basis of the temperature
dependence of the optical conductivity and in section V in terms
of persistent currents in mesoscopic rings.

\section{From local to long-range phase correlations}

The spectral features of the fermions are determined by
\begin{eqnarray}
A_F(k,\omega) &=& A^+_F(k,\omega) + A^-_F(k,-\omega) \nonumber \\
A^{\pm}_F(k,\omega)&=& \Omega^{-1}\sum_{m,n}|\langle m|{c_k^+
\choose c_k}|n\rangle|^2
e^{-\beta(E_n-\mu n_{tot})} \nonumber \\
&\*&\times\;\delta(\hbar \omega +2\mu -E_m+E_n) \nonumber \\
\Omega^{-1} &=& \sum_n e^{-\beta(E_n-\mu n_{tot})} \; ,
\end{eqnarray}
where $A^{\pm}_F(k,\omega)$ represent the electron and hole
spectral functions, respectively, $\Omega$ denotes the partition
function and $E_{m}$ are the eigenvalues of the total Hamiltonian.
In Fig.~\ref{AF-T} we plot the evolution of the fermionic spectral
function for electrons at the Fermi surface, i.e., $k=k_F$. We
notice that with decreasing temperatures spectral weight is
transferred from the Fermi energy to side bands which characterize
the contributions from bonding and antibonding two-electron
states. This emptying out of the spectral density near the Fermi
energy is responsible for the opening of a pseudogap in the DOS of
the fermions below a certain characteristic temperature $T^*$,
which can be determined from the drop in spectral weight of
$A_F(k_F,\mu)$.

The onset of local electron pairing, seen in the opening of a
pseudogap in the electronic DOS, is  manifest moreover in a strong
increase of the local correlation function between the hard-core
bosons and the electron pairs $\langle\rho^+_i c_{i\downarrow}
c_{i\uparrow} \rangle$. This effect is already inherent in a
single-site system and is illustrated in Fig.~\ref{CF}a, showing a
characteristic increase of this correlation function for
temperatures of the order of $g$, which is the energy gain of two
electrons when they engage in a bonding state $(1/\sqrt2)[\rho_i^+
+ c^+_{i\uparrow}c^+_{i\downarrow}]|0\rangle$. Going beyond the
single-site case, this energy gain is reduced to $g^2/t$ as a
consequence of the electron itinerancy entering into competition
with the charge exchange mechanism. The intensity of the on-site
correlation between bosons and electron pairs then correspondingly
decreases, as illustrated in Fig.~2b for different total numbers
of charge carriers, corresponding to $n_B= 1, 2, 3, 4$. Finally,
in Fig.~2c we examine the long-range phase coherence for the
hard-core bosons, described by $\langle\rho^+_{q=0}
\rho^-_{q=0}\rangle = (1/N)\sum_{\delta=1}^8 \langle
\rho^+_i\rho^-_{i+\delta} \rangle$. We notice that this
correlation function strengthens as $T$ drops below a certain
characteristic temperature $T_{\phi}$ ($< T^*$).

The evolution of $T^*$ and $T_{\phi}$ noticeably changes with the
concentration of bosons. We derive these temperatures from the
inflexion point of these two correlation functions and present
them as a function of $n_B$ in Fig. \ref{Tc}. While for small
boson concentration $T_{\phi}$ follows the same behavior as $T^*$,
it begins to show a downwards trend upon further increasing the
boson concentration. This confirms a recent result for the phase
diagram of the BFM obtained within a diagrammatic technique which
explicitly takes into account the hard-core nature of the bosons
\cite{Ranninger-03}. In that study it was concluded that this
non-monotonic behavior of $T_{\phi}$ should be related to the
correlation effects in the hard-core boson subsystem, leading to
an increasingly heavy mass of the bosons as their concentration
approaches the dense limit, i.e., $n_B =0.5$. This is to be
expected for systems where the onset temperature of phase ordering
is related to the phase stiffness $D_{\phi} \simeq n_p/m_p$ of the
electron pairs which is induced by the boson-fermion exchange
coupling, $n_p = \langle n_{F\uparrow}n_{F\downarrow} \rangle -
\langle n_{F\uparrow} \rangle \langle n_{F\downarrow} \rangle$
denoting their density and $m_p$ their mass.

\section{Boson spectral properties: a comparison with XY phase
fluctuation scenarios}

\begin{figure}
\begin{center}
\includegraphics[width=3.4in,height=4.4in]{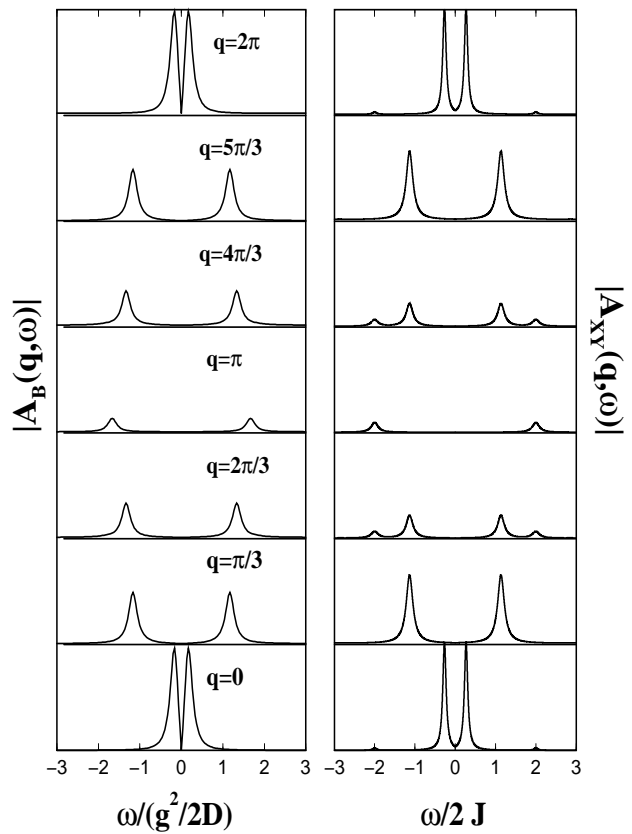}
\end{center}
\caption{\label{AB-XY} Comparison at $T=0$ of the boson spectral
function $A_B(q,\omega)$ for $n_{tot}=12$ ($n_B=3$) on a 6-site
ring with the spectral function of an equivalent XY model (denoted
by $A_{XY}(q,\omega)$).}
\end{figure}

After having considered the spectral properties of the fermions in
the preceding section, let us now discuss the spectral properties
of the hard-core bosons. The appropriate spectral function in this
case is
\begin{eqnarray}
A_B(q,\omega) &=& A^+_B(q,\omega) - A^-_B(q,\omega) \nonumber \\
A^{\pm}_B(q,\omega)&=& \Omega^{-1}\sum_{m,n}|\langle
m|\rho^{\pm}_q|n\rangle|^2
e^{-\beta(E_n-\mu n_{tot})} \nonumber \\
&\*&\times\;\delta(\hbar \omega +2\mu -E_m+E_n) \; ,
\end{eqnarray}
with $\rho_q^{\pm} = \frac{1}{\sqrt N}\sum_i
e^{iqr_i}\rho^{\pm}_i$. We ultimately want to compare this
spectral function, evaluated within the BFM, with that of an
effective XY model in a transverse field, described by
\begin{equation}
H_{XY} = -J \sum_{i,\delta} \rho^+_i\rho^-_{i+\delta} + h \sum_i
\left(\rho^z_i + \frac{1}{2}\right) \; .
\end{equation}
The corresponding results are presented in Fig.~\ref{AB-XY},
showing the absolute values of these correlation functions (we
recall that they are negative for negative frequencies $\omega$).
For the case considered here, a 6-site ring with 3 bosons and
$\Delta_B=0$, we have a chemical potential $\mu=0$ and thus we
choose $h=0$ in order to describe an equivalent physical situation
within a XY model. The typical energy scale of the bosonic
excitations is of the order of $g^2/2D$ which is the energy scale
related to the exchange interaction allowing the transfer of a
boson from a given site to an adjacent one. The strong similarity
between these two spectral functions over the entire range of $q$
vectors is a strong indication that for low temperatures ($T=0$ in
the present case) the excitation spectrum of the BFM is determined
by phase fluctuations of the bosonic charge carriers.

\begin{figure}
\begin{center}
\includegraphics[width=3.3in,height=4.4in]{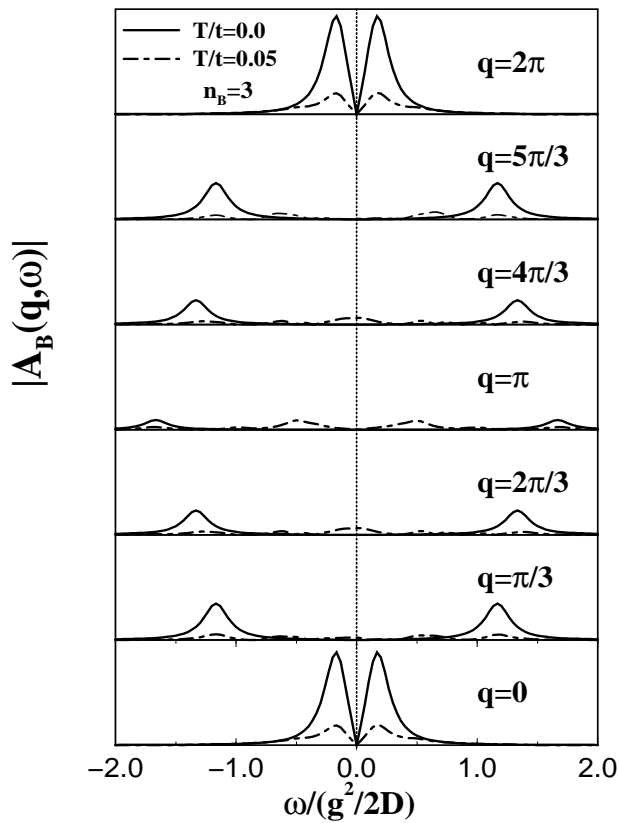}
\end{center}
\caption{\label{AB-T} Evolution with temperature of the spectral
properties of the hard-core bosons for a 6 site ring with 3 bosons
($n_{tot}=12)$ for two temperatures below and above $T_{\phi}$.}
\end{figure}

In Fig.~\ref{AB-T} we illustrate for the fully symmetric case
($\Delta_B=0$ and $n_{tot}=12$ for a 6-site ring) how the spectral
function of the hard-core bosons evolves as a function of
temperature. We notice that upon increasing the temperature above
$T_{\phi}$, this spectral function becomes more and more
incoherent. The quasi-particle features of the fermions and of the
hard-core bosons develop in opposite directions as the temperature
is lowered. Above $T^*$ the fermions are still rather well defined
quasi-particles, while the hard-core bosons are completely
incoherent. Between $T^*$ and $T_{\phi}$ the fermions lose their
quasi-particle features, while those of the bosons are getting
better and better defined. Finally, below $T_{\phi}$ the only good
quasi-particles are the hard-core bosons and consequently also the
local electron-pairs. This change in the quasi-particle features
from single to two-particle states will become more apparent in
the transport properties, discussed in the following Section.

\section{Fermion versus boson transport}

The spectral properties of the fermionic single-particle
excitations, examined in Section II, indicate a loss of spectral
weight for wave vectors centered around $k_F$, which is
responsible for the opening of the pseudogap in the DOS. Full
many-body calculations, involving self-energy corrections
\cite{Ranninger-95a,Robin-98}, have shown that besides this loss
of spectral weight, fermions also lose their quasi-particle
features below $T^*$. They become purely diffusive and show up in
a resistivity which monotonically increases as the temperature
tends to zero. At the same time, the bosons and respectively the
electron pairs, which are diffusive above $T^*$, become well
defined quasi-particles upon approaching $T_{\phi}$, as discussed
in section III. Thus, the bosonic charge carriers with charge $2e$
are expected to ensure the transport upon approaching $T_{\phi}$
and eventually lead to a drop in the resistance and ultimately to
a superconducting state.

These features can be illustrated within the present exact
diagonalization study in terms of the optical conductivity arising
from the fermions and the bosons, respectively. The electric field
in the present model couples exclusively to the fermions, since
the bosons are intrinsically localized. Nevertheless, because of
the boson-fermion exchange coupling the bosons, and consequently
also the fermion pairs, acquire itinerancy (as shown in Section
III) and thus contribute to the transport via an Aslamazov-Larkin
type contribution to the conductivity \cite{Devillard-00}. In
order to capture this effect within the present exact
diagonalization study, we decompose the current carried by the
fermions $j_F=
t\sum_{i,\delta,\sigma}(c^+_{i,\sigma}c_{i+\delta,\sigma}-h.c.)$
into a term associated with a current carried by the bosons $j_B =
t_B \sum_{i,\delta}(\rho^+_i \rho^-_{i+\delta} - h.c.)$ plus a
remainder $\widetilde{j}_F$ such that
\begin{equation}
j_F(t)=j_B(t) + \widetilde{j}_F(t) \; .
\end{equation}
Separating optimally the contributions coming from the
current-current correlation function for the bosons from the
remainder of the fermions, we require $\langle
j_B(t)\widetilde{j}_F(t)\rangle=0$, which defines an effective
boson hopping integral as $t_B = \langle
j_F(t)j_B(t)\rangle/\langle j^2_B(t)\rangle$. Such a decomposition
is meaningful in a situation where the current transported by the
fermions is rapidly dissipated as a function of time, described by
$\langle\widetilde{j}_F(t)\widetilde{j}_F(t')\rangle$, and its
long time behavior is getting controlled by the current
transported by the bosons, described by $\langle
j_B(t)j_B(t')\rangle$. Such a situation is realized for low
temperatures, around and below $T_{\phi}$. We can then safely
neglect the contributions coming from the cross terms $\langle
j_F(t)j_B(t')\rangle$.

\begin{figure}
\begin{center}
\includegraphics[width=3.4in,height=4.0in]{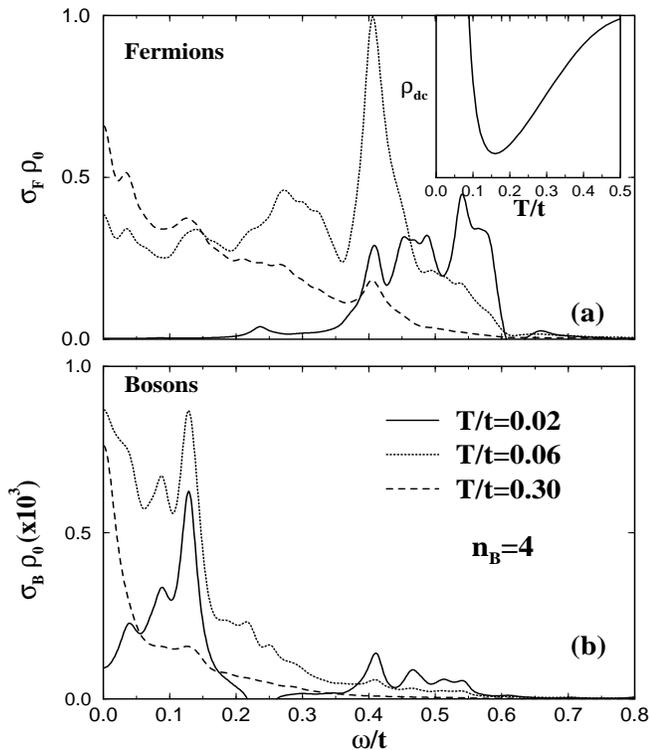}
\end{center}
\caption{\label{optcond} Optical conductivity due to fermions
(Fig.~4a) and due to bosons (Fig.~4b) for an 8-site ring with
$n_{tot}= 16$ ($n_B = 4$) and three characteristic temperatures,
corresponding to above $T^*$, between $T^*$ and $T_{\phi}$ and
below $T_{\phi}$. The displayed difference in the amplitude of the
fermionic and bosonic conductivity is set by the ratio between the
effective boson and fermion hopping ($\sim g^2/2D$). The scale is
fixed by $\rho_0=\hbar a/e^2$ where $a$ is the atomic distance.}
\end{figure}

We report in Fig.~\ref{optcond}a the optical conductivity coming
from exclusively the single-electron transport, given by $\langle
j_F(t)j_F(t')\rangle$, and in Fig.~\ref{optcond}b the contribution
coming from bosonic charge carriers. We notice that for the
fermionic contribution $\sigma_F(\omega)$ to the optical
conductivity we obtain a featureless metallic-type behavior for $T
> T^*$ ($T/t=0.3$ in our case), while upon going below $T^*$,
a pseudogap structure materializes. Upon further decreasing the
temperature below $T_{\phi}$, $\sigma_F(\omega)$ shows a well
developed gap, reflecting the onset of an insulating behavior, as
it can be inferred from the temperature behavior of the $dc$
conductivity, illustrated in the inset of Fig.~\ref{optcond}a.

At the same time, when sweeping through this same temperature
interval, we notice the build-up of a structured contribution in
the bosonic component of the optical conductivity
$\sigma_B(\omega)$, which develops precisely in the frequency
interval where the pseudogap appears in the fermionic contribution
$\sigma_F(\omega)$. These features confirm the cross-over from
fermionic to bosonic transport with decreasing temperature.

Of course, the spectral weight of the bosonic absorption is much
smaller with respect to that of the fermionic sector. The
effective boson hopping is $t_B \sim g^2/2D$, and hence the
relative spectral weight of the bosonic absorption goes like $\sim
t_B^2$. Nevertheless, it becomes clear from Fig.~\ref{optcond}
that below $T_{\phi}$ the transport at low frequency ($\omega/t <
g^2/2D$) is completely due to bosons and pairs of fermions.

\section{Persistent currents in mesoscopic rings}

The change-over between single-particle and two-particle transport
is a salient feature of the BFM and its experimental verification
would be vital in verifying such a scenario. As we have seen
above, the onset of bosonic transport is not necessarily linked to
a superconducting state. It is sufficient to have phase-coherent
free particle-like bosonic excitations. In mesoscopic rings where
persistent currents are a characteristic feature
\cite{Buttiker-83}, it should be possible to detect this
change-over in transport when threading a magnetic flux through
such a ring. This can give rise to persisting currents,
periodically varying as a function of the flux $\Phi$. In units of
the flux quantum $\Phi_0 = \hbar c/e$ this periodicity should then
be seen in multiples of $\Phi/\Phi_0= n$ if the transport is via
single electrons, and of $\Phi/\Phi_0=n/2$ if it is due to
two-particle bosonic transport, $n$ denoting an integer number.
\begin{figure}
\begin{center}
\includegraphics[width=3.4in,height=3.6in]{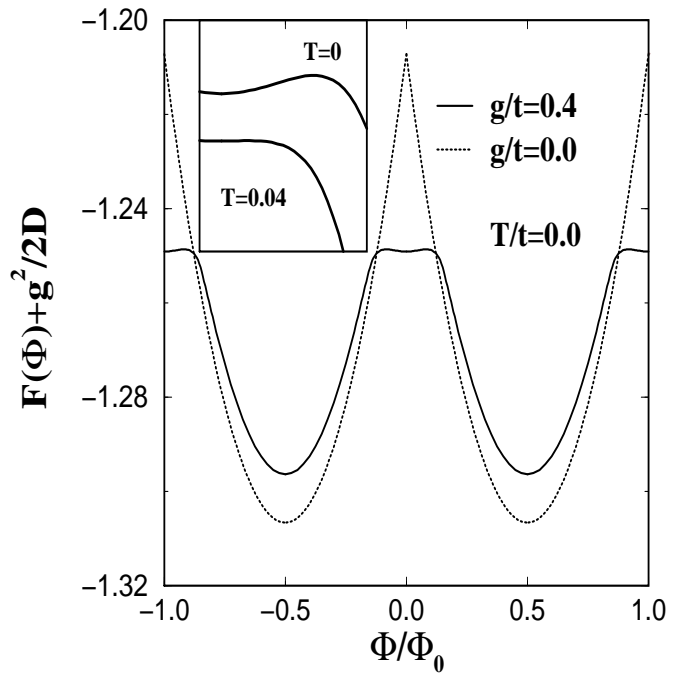}
\end{center}
\caption{\label{flux} Comparison of the free energy $F(\Phi)$ as a
function of the flux $\Phi$ for the uncoupled ($g=0$) and the
coupled ($g \neq 0$) boson-fermion system. In the inset we present
the evolution of $F(\Phi)$ for positive $\Phi$ close to 0, for
$T=0$ and $T=0.04$ ($> T_{\phi}$)}
\end{figure}

\begin{figure}
\begin{center}
\includegraphics[width=3.4in,height=3.6in]{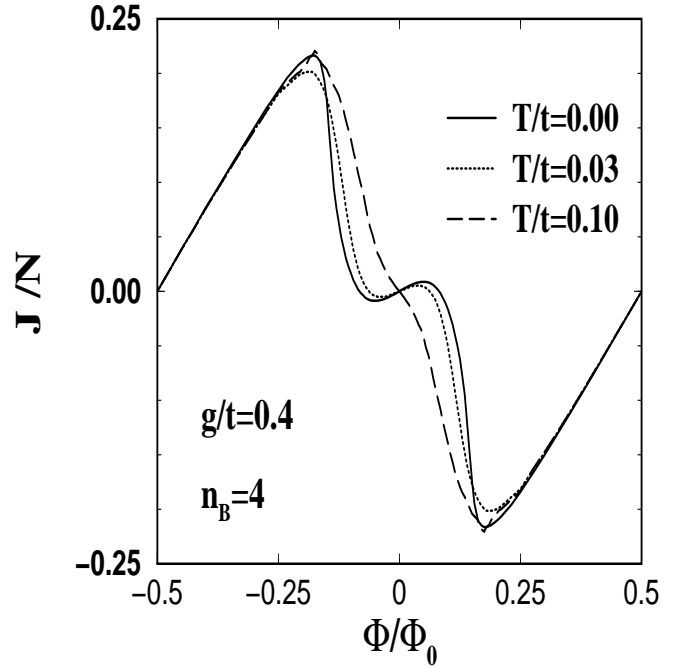}
\end{center}
\caption{\label{flux1} Evolution with temperature of the current induced
by a magnetic field threading a mesoscopic ring, illustrating the change
in periodicity of the current as we pass from temperatures above $T_{\phi}$
to below it.}
\end{figure}
We investigate this problem in the standard way by evaluating the
free energy $F(\Phi)$ in the presence of a magnetic field and then
determine the resulting current as $J/N=\partial F(\Phi)/\partial
\Phi$. The behavior of $F(\Phi)$ follows from the evaluation of
the thermal average of the Hamiltonian (1), upon making the
Peierls substitution $t \rightarrow
t\,\exp\left(-i\frac{\Phi}{N\Phi_0}\right)$. We illustrate in
Fig.~\ref{flux} the variation in the ground state of $F(\Phi)$ as
a function of $\Phi/\Phi_0$ for the uncoupled system ($g=0$) and
the fully interacting system ($g\ne 0$). In the inset of the
figure we also report the behavior of $F(\Phi)$ for small positive
values of $\Phi$ at $T=0$ and at a finite temperature
$T>T_{\phi}$, in the case $g/t=0.4$. As to be expected, for the
case $g=0$ we find a situation where the applied field lowers the
energy of the system, arriving at minima at $\Phi/\Phi_0=n/2$ (see
dotted line in Fig.~\ref{flux}). The periodicity of one flux
quantum of $F(\Phi)$ suggests that the effect of the flux is to
introduce a paramagnetic current made out of single-electron
states. Upon switching on the boson-fermion interaction, $g\neq
0$, one finds that around $\Phi = 0$ the free energy below
$T_{\phi}$ increases with increasing $\Phi$ before bending over
and following essentially the same behavior as that obtained for
$g=0$. This suggests that if a small flux is applied to the
system, a diamagnetic screening current is building up which,
eventually, gives way to a paramagnetic current stabilizing the
system when the flux is further increased. Hence, upon switching
on the boson-fermion interaction, the system evolves towards a
state where the free energy exhibits minima at values of the flux
with periodicity $\Phi_0 / 2$. This is indicative of an induced
current made out of carriers having charge $2e$, and of an
off-diagonal long-range order in the ground state. These
conclusions are based on results derived a long time ago
\cite{Byers-61},\cite{Yang-62} in order to provide a criterion to
distinguish between normal and superconducting ground states from
the functional form of $F(\Phi)$, without any knowledge of the
symmetry of pair-pair correlation functions. The fact that the
minimum of the free energy at $\Phi/\Phi_0=0$ is not the same as
at $\Phi/\Phi_0=1/2$ is due to the finite-size calculation
presented here, analogous to the study of the anomalous flux
quantization in the negative $U$ Hubbard problem (see Fig.~1 in
ref. \cite{Stafford-93}). As the temperature is increased above
$T_{\phi}$, we notice that the minimum of $F$ at $\Phi=0$
gradually disappears, thus stabilizing the paramagnetic
single-electron current.

In Fig.~\ref{flux1} we present the current induced by the magnetic
flux threading an 8-site mesoscopic ring as a function of
$\Phi/\Phi_0$, for three characteristic temperatures, two below
and one above $T_{\phi}$. We again can see the change-over from
two-particle to single-particle transport. Experiments on
mesoscopic ring made out of a high-$T_c$ compound might possibly
have seen such two-particle transport at low temperatures
\cite{Kawabata-98}.

\section{Conclusions}

The present exact diagonalization study of the BFM confirms
previous results obtained by various many-body techniques, all of
which have their built-in shortcomings. The competitive effect of
local and long-range correlations is a new element in the present
approach, allowing to extract some relevant physics underlying the
BFM . They are characterized by the energy scales $k_B T^*$ and
$k_B T_{\phi}$, describing respectively the onset of local pairing
and the establishment of a phase-correlated state on a long range,
the latter being indicative of a superconducting phase in an
infinite system. The change-over from single-particle to
two-particle transport is most clearly seen and discussed in terms
of the optical conductivity, showing below $T^*$ a single-particle
pseudogap feature together with a two-particle contribution
developing in the frequency window opened up by the pseudogap. A
clear-cut experiment for verifying the present scenario would be
to study the change-over form single-electron to two-electron
transport in mesoscopic rings, the theoretical expectations for
which have been discussed.

In the present study, we limited the analysis to the case of
optimal resonant scattering between the electrons and the local
bound electron pairs, this being the physically most interesting case
for the interplay between phase and amplitude fluctuations. Such a
situation is realized when the bosonic level $\Delta_B$ lies in
the middle of the fermionic band (i.e. $\Delta_B=0$). By changing
the position of the bosonic level, there are no qualitative
changes as far as concerns the static and dynamical response and
the appearance of two characteristic energy scales which control
the short and long range phase coherence. In particular, the three-
peak structure in the density of states (Fig. 1) remains the same,
but the spectral weight is now asymmetric with respect to the
chemical potential.

\section*{ACKNOWLEDGMENTS} M.C. acknowledges support from Marie
Curie Fellowship within the Program "Improving Human Potential".

\bibliography{art}

\end{document}